\documentclass[pre,tighten,twocolumn]{revtex4}

\usepackage{graphicx}

\begin{document}

\hyphenation{nano-tube nano-tubes}

\title{Biasing the center of charge in molecular dynamics simulations
with empirical valence bond models: free energetics of an excess proton in a water droplet}

\author{J\"urgen K\"ofinger and Christoph Dellago}
\address{Faculty of Physics and Center for Computational Materials Science,
University of Vienna,                    
Boltzmanngasse 5, 1090 Vienna, Austria}

\date{\today}

%-------------------------------------------------------------------------
%  ABSTRACT
%-------------------------------------------------------------------------
\vspace*{1cm}
\begin{abstract}
Multistate empirical valence bond (EVB)  models provide an accurate description 
of the energetics of proton transfer and solvation in complex molecular 
systems and can be efficiently used in molecular dynamics computer simulations. 
Within such models, the location of the moving protonic charge can be 
specified by the so called center of charge, defined as a weighted 
average over the diabatic states of the EVB model. In this 
paper, we use first order perturbation theory to calculate
the molecular forces that arise if a bias potential is applied to the center 
of charge. Such bias potentials are often necessary when molecular dynamics
simulations are used to determine free energies related to proton transfer and 
not all relevant proton positions are sampled with sufficient frequency during 
the available computing time. The force expressions we derive are easy to 
evaluate and do not create any significant computational cost compared with
unbiased EVB-simulations. As an illustration of the method, we study proton 
transfer in a small liquid water droplet consisting of 128 water molecules plus 
an excess proton. Contrary to predictions of continuum electrostatics but 
in agreement with previous computer simulations of similar systems, we observe that 
the excess proton is predominantly located at the surface of the droplet. Using 
the formalism developed in this paper, we calculate the reversible work 
required to carry the protonic charge from the droplet surface 
to its core finding a value of roughly 4 $k_{\rm B}T$.
    
\end{abstract}

\maketitle

%-------------------------------------------------------------------------
%  INTRODUCTION
%-------------------------------------------------------------------------
\section{Introduction}
\label{sec:introduction}

Proton transfer is of crucial importance for a variety of processes in nature and technology ranging from ATP synthesis in living cells \cite{WIKSTROEM} and enzymatic catalysis \cite{SILVERMAN} to  chlorine chemistry on stratospheric ice particles involved in polar ozone depletion \cite{MOLINA,BUCH,HYNES} and electrical power generation in hydrogen fuel cells \cite{KREUER,COLOMBAN}. As proton transfer involves cleavage and formation of covalent bonds its computer simulation is challenging. Ab initio methods such as density functional theory or wavefunction based methods can model chemical reactivity but are computationally extremely expensive \cite{MARX,HUTTER}. Molecular dynamics trajectories obtained with such methods are short and the number of collected transfer events is not sufficient to carry out a thorough statistical analysis capable of revealing the details of the mechanism. Recently, however, computationally efficient empirical valence bond (EVB) models for proton transport have been developed by Voth and collaborators \cite{VOTH1,VOTH2,CUMA1,CUMA2,VOTH_REVIEW} and Vuilleumier and Borgis \cite{BORGIS1,BORGIS2,BORGIS3} based on pioneering work of Warshel \cite{WARSHEL1,WARSHEL2,WARSHEL3}. These models, in which the Born-Oppenheimer surface is obtained as the lowest root of a secular equation involving empirically modeled diabatic states and coupling elements, accurately describe the energetics of bond cleavage and formation but are computationally far less expensive than ab initio methodologies. With forces computed with the Hellmann-Feynman theorem one can perform nanosecond molecular dynamics simulations during which many proton transfer events occur. 

Empirical valence bond models have been used to study proton solvation and transport in many systems including bulk water \cite{BORGIS3,VOTH2}, water clusters \cite{EVB_TRIMER,VOTH_CLUSTER1,VOTH_CLUSTER2}, water filled pores \cite{DellagoPRL,PRL_MEMBRANE,VOTH_PORE,VOTH_SCHULTEN}, acidic aqueous solutions \cite{CUMA1,CUMA2}, the water-vapor interface \cite{WATER_VAPOR}, and biological systems \cite{WARSHEL1,WARSHEL2,WARSHEL3,VOTH_BIO1,VOTH_BIO2}. Many of these applications have been surveyed in a recent review article by Voth and collaborators \cite{VOTH_REVIEW}. A quantitative understanding  of proton transport and solvation often requires the calculation of free energies related to different positions of the excess proton. For instance, the rate for proton translocation across membrane nanopores is mainly determined by the free energetic cost required to remove the proton from the bulk and bring it into the interior of the pore \cite{PRL_MEMBRANE}. The definition of the position of the excess proton, however, is not straightforward in empirical valence bond models. In bulk water, for instance, the excess proton occurs in a variety of different configurations with the so called Eigen and Zundel cations as the limiting cases  \cite{TUCK,MARX,VOTH2}. Whereas in the Eigen cation, ({H$_9$O$_4$})$^+$, the proton is strongly associated with one particular water molecule that donates three contracted hydrogen bonds to adjacent water molecules, the excess charge is equally shared by two water molecules in the Zundel cation,  ({H$_5$O$_2$})$^+$. Under the influence of the fluctuating hydrogen bond network, these structures freely convert into each other by proton hops along hydrogen bonds on a picosecond time scale. Due to this fluxional character of the hydrated proton, identification of a particular proton as the excess charge is ambiguous.

In the empirical valence bond model this ambiguity can be resolved by defining the {\em center of charge}, the weighted average of the positions of the excess charge in the diabatic states \cite{EVB_NEW}. The center of charge provides a meaningful definition of the proton position in a way that changes continuously in time. The free energetics of proton translocation can then be determined in molecular dynamics simulations by accumulating statistics on the position of the center of charge. This straightforward approach works well if the free energy differences do not exceed the thermal energy $k_{\rm B}T$ appreciably. If they do, the molecular dynamics simulation most likely fails to explore all important regions of configuration space within the available computing time. In such cases, it is possible to enhance the sampling of configuration space by introducing appropriate bias potentials that guide the system towards configurations that would not be sampled otherwise \cite{FRENKEL_SMIT}. For instance, in the case of proton transfer through a membrane, penetration of the proton into the pore interior is observed in a molecular dynamics simulation only if the related desolvation penalty is compensated by a suitable bias. Correction for the bias then permits to deduce ensemble averages and free energies for the bias-free system.

When bias potentials are used in molecular dynamics simulations it is necessary to calculate the atomic forces resulting from the bias. In the case of a bias potential acting on the center of charge, such a calculation requires to determine the derivatives of the center of charge with respect to the particle coordinates. How to do that using first order perturbation theory is the central subject of this paper. The expressions obtained in this way permit a computationally inexpensive evaluation of the bias forces because they depend only on quantities already required in the calculation of regular EVB-forces. As an example, we apply the formalism developed in this way to study the solvation of an excess proton in a small water droplet. We note that free energies as a function of the position of the center of charge have been calculated before in biased EVB-simulations \cite{VOTH_SCHULTEN,VOTH_AQUAPORIN1,VOTH_AQUAPORIN2}. In these publications, however, the calculation of the forces resulting from the bias is not explained and it is unclear how it is done. 

The remainder of this article is organized as follows. In the next section we will briefly review the multi-state empirical valence bond model and define the center of charge. First order perturbation theory is then used in Sec. \ref{sec:bias} to derive the atomic forces resulting from a bias potential acting on the center of charge. To illustrate the formalism, we calculate the reversible work required to transfer an excess proton from the surface to the interior of a small water droplet defined in Sec. \ref{sec:system} and present the results in \ref{sec:results}. Some conclusions are given in Sec. \ref{sec:conclusions}.    

%-------------------------------------------------------------------------
%  EVB
%-------------------------------------------------------------------------
\section{Multi-state empirical valence bond model}
\label{sec:evb}

In this section we briefly review the multi-state empirical valence bond model in order to introduce the terminology and set the notation. For a more detailed description we refer the reader to the original publications \cite{VOTH1,BORGIS1}. The system, which includes of a certain number of water molecules plus an excess proton and possibly some other species, is described by $3N$ atomic coordinates $x=\{x^1,x^2, \cdots, x^{3N}\} $. The position of particle $i$ is also denoted by ${\bf r}_i$ such that we can write $x=\{{\bf r}_1, {\bf r}_2, \cdots,  {\bf r}_N\}$ when it is convenient. In the empirical valence bond model one imagines that the electronic state $| \psi\rangle$  for a particular atomic configuration is a superposition of $L$ {\em diabatic} valence bond states $|\varphi_i\rangle$,
\begin{equation}
\label{equ:superposition}
| \psi\rangle = \sum_{i} c_i |\varphi_i\rangle,
\end{equation}
where the $c_i$ are real expansion coefficients. In each state chemical bonds are assigned such that a different oxygen atom holds the hydronium ion H$_3$O$^+$. In some states this assignment may correspond to a rather distorted geometry of the hydronium ion and the neighboring water molecules. Such states will have a high energy and therefore contribute only little to the ground state. Only a small number of states are considered which are chosen to account for different plausible routes for proton transfer. To construct these states, one starts from a pivot state and then includes other states that are accessible \cite{VOTH1}. Typically, about 10 states are required to accurately model proton transfer in bulk water. 

To find the Born-Oppenheimer energy surface as a function of the atomic coordinates one does not determine the electronic state $| \psi\rangle$ from first principles. Rather, all $L^2$ matrix elements $H_{ij}(x)=\langle\varphi_i| H|\varphi_i\rangle$ of the electronic Hamiltonian $H$ are modeled empirically as functions of the atomic coordinates $x$. While the $L$ diagonal elements are the potential energies corresponding to particular hydronium ions with fixed chemical bonding topology, the $L(L-1)/2$ non-diagonal elements provide the coupling between the diabatic states that enables chemical reactivity. Postulating that the diabatic states are orthogonal to each other, solution of the eigenvalue problem
\begin{equation}
H {\bf c}_i = E_i {\bf c}_i
\end{equation}
yields the adiabatic ground state. Here, ${\bf c}_i\equiv\{c_{i0}, c_{i1}, \cdots, c_{i(L-1)}\}$ is the eigenvector of dimension $L$ belonging to the eigenvalue $E_i$ and its elements are the corresponding coefficients in the superposition of Equ. (\ref{equ:superposition}). (Here and in the following we number states starting with $0$.) The Born-Oppenheimer potential energy surface is then given as the expectation value of the energy in the ground state, 
\begin{equation}
E_0(x)=\sum_{i,j} c_{0i}c_{0j}H_{ij}(x),
\end{equation}
where $c_{0i}$ is the coefficient of state $i$ in the ground state. The corresponding forces can be calculated using the Hellmann-Feynman theorem \cite{HELLMANN_FEYNMAN}:
\begin{equation}
\label{equ:HF}
F^\alpha = -\sum_{i,j} c_{0i}c_{0j}\frac{\partial H_{ij}(x)}{\partial x^\alpha}.
\end{equation}
Here, $F^\alpha$ is the force acting on degree of freedom $\alpha$. This procedure, which retains quantum mechanics on a rudimentary level, effectively interpolates in a smart way between the diabatic states for which empirical potential energy surfaces for fixed chemical bonding topology  are known.

We can now define the {\em center of charge} ${\bf q}_c$ as the weighted average of the hydronium oxygen position over all diabatic states,
\begin{equation}
{\bf q}_c=\sum_{i} c_{0i}^2 {\bf q}_i.
\label{equ:coc}
\end{equation}
Here, ${\bf q}_i$ is the position of the hydronium oxygen in diabatic state $i$ and $c_{0i}^2$ is the statistical weight of the diabatic state $i$ in the adiabatic ground state. This definition of the center of charge takes into account the delocalized nature of the protonic charge and can be used to follow its migration along the fluctuating hydrogen bond network.

%-------------------------------------------------------------------------
%  BIAS FORCES
%-------------------------------------------------------------------------

\section{Bias Forces}
\label{sec:bias}

We now imagine that a bias potential $V_{\rm B}({\bf q}_c)$, 
which depends on the center of charge ${\bf q}_c$, is added to the
potential energy of the system. (If the bias potential is a function 
of other degrees of freedom in addition to ${\bf q}_c$, the formalism 
below needs to be adapted appropriately.) Depending on the particular
situation one wants to study, the bias potential $V_{\rm B}({\bf q}_c)$
is designed to control the position of the 
protonic charge and enhance the sampling of the configuration 
space regions of interest. If one uses such a function in a molecular dynamics
simulation it is necessary to determine the corresponding forces via
\begin{equation}
F_{\rm B}^\alpha(x)=-\frac{\partial V_{\rm B}({\bf q}_c)}{\partial
x^\alpha}.
\label{equ:force}
\end{equation}
In the following we will derive an explicit expression that permits 
to calculate exactly these forces.

Using the chain rule and the definition of the center of charge from 
Equ. (\ref{equ:coc} )we
can write the derivative of $V_{\rm B}({\bf q}_c)$ with respect to the
coordinate $x^\alpha$ as
\begin{equation}
\label{equ:dVdx} \frac{\partial V_{\rm B}({\bf q}_c)}{\partial
x^\alpha}=  \sum_{\beta=1}^{3} \frac{\partial V_{\rm B}({\bf
q}_c)}{\partial
q^\beta_c}\frac{\partial q^\beta_c}{\partial x^\alpha},
\end{equation}
where $q^\beta_c$ is coordinate $\beta$ of the center of charge ${\bf
q}_c$. To calculate the derivatives $\partial q^\beta_c / \partial x^\alpha$ of the center or charge with respect to the particle coordinate we differentiate Equ. (\ref{equ:coc}).  Taking into account that also the ground state coefficients $c_{0i}$ depend on the particle coordinates we obtain
\begin{equation}
\label{equ:dqdx}
\frac{\partial q^\beta_c}{\partial x^\alpha}= 2 \sum_{i}  c_{0i} \frac{\partial
c_{0i}}{\partial x^\alpha}q_i^\beta + \sum_i c_{0i}^2 \frac{\partial
q_i^\beta}{\partial x^\alpha}.
\end{equation}
Here, $q^{\beta}_i$ is coordinate $\beta$ of the hydronium oxygen
of diabatic state $i$. The partial derivative $\partial
q_i^\beta/\partial x^\alpha$ is unity only if $x^\alpha$ is the
coordinate $\beta$ of the oxygen carrying the hydronium in state $i$
and it vanishes otherwise.

According to Equ. (\ref{equ:dqdx}), we need to evaluate derivatives of
the elements of the ground state eigenvector ${\bf c}_0$ with respect to the
coordinates of all particles. This can be accomplished with ordinary
first order perturbation theory (see, for instance, Ref. \cite{QM}). 
To do so, we first consider how the
matrix element $H_{ij}$ of the Hamiltonian changes if one particular coordinate
$x^\alpha$ is changed by an infinitesimal amount $\lambda$, i.e.,
\begin{equation}
H_{ij} \rightarrow H_{ij} + \lambda \frac{\partial H_{ij}}{\partial x^\alpha }=H_{ij} + \lambda D_{ij}^\alpha,
\end{equation}
where we have introduced
\begin{equation}
D^{\alpha}_{ij}\equiv\frac{\partial}{\partial
x^\alpha}H_{ij},
\end{equation}
Note that the derivatives $D^{\alpha}_{ij}$ are already 
available from the regular EVB force calculation. 

To calculate the derivative of $c_{0i}$ with respect to
$x^\alpha$ we need to determine how $c_{0i}$ changes if we add the
perturbation $\lambda D^\alpha$ to the Hamiltonian $H$. (Here, $D^\alpha$ denotes the matrix consisting of the elements $D^{\alpha}_{ij}$.) Then, we can
calculate the derivative from:
\begin{equation}
\label{equ:derivative}
\frac{\partial c_{0i}}{\partial
x^\alpha}=\lim_{\lambda \to 0}\frac{c_{0i}(\lambda)-c_{0i}}{\lambda}.
\end{equation}
The argument in $c_{0i}(\lambda)$ indicates the dependence of
$c_{0i}$ on the strength of the perturbation, i.e., on the displacement
of coordinate $x^\alpha$. If no argument is written for $c_{0i}$ the value
for a vanishing perturbation strength is implied. 

The ground state coefficients $c_{0i}(\lambda)$ as a function of
the displacement $\lambda$ can be calculated by expanding the 
coefficients and the energies in powers of $\lambda$ and truncating
the expansion after first order,
\begin{eqnarray}
E_k(\lambda)&=&E_k^{(0)}+\lambda E_k^{(1)}+O(\lambda^2),\\
c_{ki}(\lambda)&=&A(\lambda)\left\{c_{ki}+\lambda \sum_{j\neq
k}a_{kj}^{(1)}c_{ji}+O(\lambda^2)\right\},
\end{eqnarray}
where the superscript denotes the order of the term and
$A(\lambda)$ is a normalization constant. Standard time independent
perturbation theory then yields \cite{QM}
\begin{eqnarray}
\lambda E_k^{(1)}&=& \sum_{l, m}c_{kl} \lambda D^{\alpha}_{lm}c_{km},\\
\lambda a_{kj}^{(1)}&=&\frac{\sum_{l, m}c_{jl} \lambda
D^{\alpha}_{lm}c_{km}}{E_k^{(0)}-E_j^{(0)}}.
\end{eqnarray}
Since the normalization factor $A(\lambda)$ is unity to first order,
we obtain
\begin{eqnarray}
\lambda E_k(\lambda)&=& E_k^{(0)} + \lambda\sum_{l, m}D^{\alpha}_{lm}c_{kl}c_{km},\\
c_{ki}(\lambda) &=& c_{ki}+\lambda \sum_{j\neq k} \frac{\sum_{l, m}
D^{\alpha}_{lm}c_{jl}c_{km}}{E_k^{(0)}-E_j^{(0)}}c_{ji}. \label{equ:coeff}
\end{eqnarray}
Inserting Equ. (\ref{equ:coeff}) into Equ.
(\ref{equ:derivative}) we obtain
\begin{equation}
\label{equ:bias} 
\frac{\partial c_{0i}}{\partial
x^\alpha}=\sum_{j\neq 0}\sum_{l, m} \frac{
D^{\alpha}_{lm}c_{jl}c_{0m}}{E_0-E_j}c_{ji},
\end{equation}
where, for simplicity, we have omitted the superscripts for the unperturbed energies. 
Thus, we now have all elements in place necessary to calculate the atomic 
forces resulting from the bias potential using Equs. (\ref{equ:dVdx}) and (\ref{equ:dqdx}). 

When evaluating the bias forces according to Equ. (\ref{equ:dVdx}) in a simulation it is important to note that derivatives $\partial c_{0i}/\partial
x^\alpha$ of the ground state coefficients appear only in contraction with the vector $c_{0i}q_i^\beta$ (see Equ. (\ref{equ:dqdx})). Inserting Equ. (\ref{equ:bias}) into Equ. (\ref{equ:dqdx}) we obtain
\begin{equation}
\label{equ:dqdx2}
\frac{\partial q^\beta_c}{\partial x^\alpha}= 2 \Gamma^{\alpha \beta} + \sum_i c_{0i}^2 \frac{\partial
q_i^\beta}{\partial x^\alpha},
\end{equation}
where for convenience $\Gamma^{\alpha \beta}$ is defined as the multiple sum
\begin{equation}
\label{equ:Gamma}
\Gamma^{\alpha \beta}= \sum_{i}\sum_{j\neq 0}\sum_{l, m} \frac{
D^{\alpha}_{lm}c_{0m}c_{jl}c_{ji}  c_{0i} q_i^\beta}{E_0-E_j}.
\end{equation}
The quadruple summation in the above equation is the compuationally most expensive part of the calculation of the bias forces (compared to the overall cost of an EVB simulation, however, the cost of the bias force calculation is negligible). 

The number of operations required for the calculation of $\Gamma^{\alpha \beta}$ now depends on the order in which the summations are carried out. This situation is similar to that often encountered in electronic structure calculations (see, for instance, Ref. \cite{HANDY}). One may, for instance, decide to carry out the summation over the indices $m$ and $l$ first for all $j$,
\begin{equation}
B^\alpha_j\equiv \sum_{l, m} D^{\alpha}_{lm}c_{0m}c_{jl},
\end{equation}
and do summation over $j$ for all $i$ after that,
\begin{equation}
\frac{\partial c_{0i}}{\partial x^\alpha}=\sum_{j\neq
0}\frac{B_j^\alpha c_{ji}}{E_0-E_j}.
\end{equation}
Finally, $\Gamma^{\alpha \beta}$ is calculated by summing over $i$,
\begin{equation}
\Gamma^{\alpha \beta}=\sum_{i}c_{0i} q_i^\beta \frac{\partial c_{0i}}{\partial x^\alpha}.
\end{equation}
The total number of operations of this procedure is of order $L^3$ (recall that $L$ is the number of diabatic states).

While most summation orders lead to the same $L^3$-scaling, there is a particular way to carry out the summations for which the number of required operations scales as $L^2$. This is achieved by first summing over $i$ for all $j$,
\begin{equation}
Q_j^{\beta} = \sum_i c_{ji}  c_{0i} q_i^\beta.
\end{equation}
Then, one carries out a summation over $j$ for all $l$,
\begin{equation}
U_l^{\beta} = \sum_{j\neq 0}\frac{c_{jl} Q_j^\beta }{E_0-E_j}.
\end{equation}
The last step consists of a summation over $m$ and $l$,
\begin{equation}
\Gamma^{\alpha \beta} =  \sum_{l, m} D^{\alpha}_{lm} c_{0m} U_j^\beta.
\end{equation}
In this case, each of the three steps requires of the order of $L^2$ operations so that also the total number of operations is of that order.

Using these expressions one can calculate the forces caused
by the bias potential at little extra cost and use them in a
molecular dynamics simulation. Due to the denominator containing the
energy difference between the ground state energy and higher
eigen-energies of the EVB-Hamiltonian the above expressions are valid
only in the non-degenerate case. Although the degenerate case can be
treated with slightly modified expressions, this has never been
necessary in our simulations.

We close this section on the formalism by mentioning an alternative viewpoint that, however, yields exactly the same expressions for the bias force. This perspective is based on the observation that the center of charge can be expressed as a derivative of the (appropriately modified) total energy with respect to the components of constant electric field ${\cal E}=\{{\cal E}^1, {\cal E}^2, {\cal E}^3\}$ which couples to the hydronium oxygens of the diabatic states. More specifically, the elements of the Hamiltonian matrix are modified in the following way:
\begin{equation}
H_{ij}(x, {\cal E})=H_{ij}(x)+\delta_{ij} ({\cal E}\cdot {\bf q}_i).
\end{equation}
Then, the Hellmann-Feynman theorem can be used to show that the components of the center of charge are given by derivatives with respect to the field,
\begin{equation}
q_c^\beta = \frac{\partial E_0(x, {\cal E})}{\partial {\cal E}^\beta}.
\end{equation}
Using this relation, the derivatives of the center of charge with respect to the particle coordinates can be expressed as 
\begin{equation}
\frac{\partial q_c^\beta}{\partial x^\alpha}=\frac{\partial^2 E_0(x,{\cal E})}{\partial x^\alpha\partial {\cal E}^\beta}=\frac{\partial^2 E_0(x,{\cal E})}{\partial {\cal E}^\beta\partial x^\alpha}=-\frac{\partial F^\alpha}{\partial {\cal E}^\beta},
\end{equation}
where we have exploited that the differentiation order can be exchanged and the derivative with respect to ${\cal E}^\beta$ is evaluated at ${\cal E}=0$. The derivatives of the forces $F^\alpha$ with respect to the particle coordinates required in the above equation can be computed from Equ. (\ref{equ:HF}) using perturbation theory. In contrast to the development described above, the perturbation is done with respect to the $3$ electric field components rather than to the $3N$ particle coordinates. Nevertheless, this approach results in exactly the expressions of Equs. (\ref{equ:dqdx}), (\ref{equ:dqdx2}), and (\ref{equ:Gamma}), which can be evaluated efficiently as explained above.   

%-------------------------------------------------------------------------
%  SYSTEM AND SIMULATIONS
%-------------------------------------------------------------------------

\section{System and Simulations}
\label{sec:system}

As an illustration of the method described in 
the previous section, we apply it to a system consisting of a cluster of $N_W=128$ water molecules plus one excess proton. Thus, the total number of atoms is $N=385$.
The interaction energies and forces were
calculated using the multi-state empirical valence bond model of Voth 
and collaborators \cite{VOTH1,VOTH2}, which was demonstrated to accurately 
describe proton transfer in bulk water. The simulations were carried 
at a temperature of $T=280$ K controlled using an Andersen 
thermostat \cite{ANDERSEN} in which particle momenta are randomly 
assigned from a Maxwell-Boltzmann distribution with a rate of 
$\nu=0.1\, {\rm ps}^{-1}$. The equations of motion were integrated using 
the velocity Verlet algorithm with a time step of $\Delta t = 0.5 $ fs for a 
hydrogen mass of 2 amu and an oxygen mass of 16 amu (a hydrogen mass of 2 rather than 1 amu leaves the structural properties of the system unchanged, but allows for a larger time step). At each time step, the 
diabatic states are determined using the algorithm developed by Schmitt 
and Voth \cite {VOTH1,VOTH2} such that the set of states is continuously adapted 
according to the location of the excess charge. The EVB-Hamiltonian is diagonalized 
using the Jacobi method for symmetric matrices. Typically, the number of 
diabatic states of the empirical valence bond model fluctuates between
8 and 10.

To prevent single water 
molecules to evaporate from the cluster we have enclosed the whole 
system in a confining potential of the form
\begin{equation}
V_{\rm conf}(x)=\frac{k_{\rm conf}}{2}\sum_i^N (r^{\rm cm}_{i}-R_{\rm conf})^2\theta(r^{\rm cm}_{i}-R_{\rm conf}),
\end{equation} 
where $\theta(x)$ is the Heaviside step function and
\begin{equation} 
r^{\rm cm}_{i}=|\mathbf{r}_i- \mathbf{R}_{\rm cm}|
\end{equation}
is the distance of atom $i$ from the center of mass $\mathbf{R}_{\rm cm}$ of the cluster,
\begin{equation}
\mathbf{R}_{\rm cm} = \frac{\sum_i^N m_i \mathbf{r}_i}{\sum_i^N m_i}.
\end{equation}
The force constant was set to $k_{\rm conf}=10$ kcal/mol \AA$^{-2}$  
and the distance 
$R_{\rm conf}$ at which the atoms start to feel the confining potential was 
set to $R_{\rm conf}=18$ {\AA}, sufficiently large to accommodate typical shape 
fluctuations of the cluster. Due to this confining potential the cluster is 
effectively in equilibrium with the vapor phase.

For the calculation of the free energy $F(r)$ as a function of the distance of
the center of charge from the center of mass we have performed a series 
of calculation in which a parabolic biasing potential
\begin{equation}
\label{equ:Ubias}
V_{\rm B}(x)=\frac{k_{\rm bias}}{2} (r-R_{\rm bias})^2
\end{equation} 
was added to the potential energy of the system. The biasing potential depends
on the configuration of the system only through
\begin{equation}
r = |\mathbf{q}_c-\mathbf{R}_{\rm cm}|,
\end{equation}
the distance of the center of charge from the center of mass.
Forces resulting from the bias potential given in Equ. (\ref{equ:Ubias})
can be calculated using the formalism presented in Sec. \ref{sec:bias}.
With a sufficiently large value of the 
force constant $k_{\rm bias}$, a biasing potential of this form forces the 
distance $r$ between the center or charge and 
the center of mass to stay close to $R_{\rm bias}$. By combining several simulations
with appropriately selected values of the parameters  $k_{\rm bias}$ and 
$R_{\rm bias}$ the complete range of interest for the distance $r$ can be sampled.
In particular, the bias potential coerces the simulation to visit configuration 
space regions that would not be visited without bias on time scales accessible 
to a simulation without bias. The simulation parameters used in this work
are listed in Table \ref{tab:simpar}.

\begin{table}[t]
\caption{\label{tab:simpar} Parameters for the runs with biasing potential.}
\begin{ruledtabular}
\begin{tabular}{rrr}
$k_{\rm bias}$ \hspace*{1cm}& $R_{\rm bias}$ & length
\\ \hline
10.0 kcal/mol \AA$^{-2}$ & 0.0 \AA & 5 ns \\
6.0 kcal/mol \AA$^{-2}$ & 1.5 \AA & 5 ns \\
6.0 kcal/mol \AA$^{-2}$ & 3.0 \AA & 5 ns \\
3.0 kcal/mol \AA$^{-2}$ & 4.0 \AA & 5 ns
\\
\end{tabular}
\end{ruledtabular}
\end{table}

As one is usually interested in the properties of the system without the bias, 
one must correct for its effect on the observables \cite{FRENKEL_SMIT}. Here we
are interested in the distribution $P(r)$ of the distance $r$, 
\begin{equation}
P(r)=\langle \delta [r-r(x)]\rangle, 
\end{equation}   
where the angled brackets $\langle \cdots \rangle$ denotes a canonical average. 
In this case, the correction for the bias yields
\begin{equation}
\label{equ:Pbias}
P(r)\propto \exp[V_{\rm B} (r)/k_{\rm B}T]\langle \delta [r-r(x)]\rangle_{\rm bias}, 
\end{equation}   
where $k_{\rm B}$ is Boltzmann's constant and  $\langle \cdots \rangle_{\rm bias}$ denotes an average in the biased ensemble. The distribution $\langle \delta [r-r(x)]\rangle_{\rm bias}$ of $r$ can be calculated 
by histogramming $r$ in the simulation with bias. In principle, the complete distribution of $r$ can be determined according to Equ. (\ref{equ:Pbias}) from one single simulation. In practice, however, the distribution $P(r)$ calculated in this manner is accurate only in the range of $r$ where in the biased simulation sufficient statistics is accumulated, i.e., in the range near $R_{\rm bias}$. To determine  the distribution $P(r)$ over a wider range of distances $r$ one has to combine distributions obtained from various simulations with appropriate values of $R_{\rm bias}$ and $k_{\rm bias}$. Systematic procedures have been developed for this purpose \cite{FERRENBERG}. Here, however, it proved sufficient to match the distributions obtained from the various biased simulations by hand by multiplication of the individual distributions with appropriate constant factors. 

%-------------------------------------------------------------------------
%  RESULTS
%-------------------------------------------------------------------------

\section{Results}
\label{sec:results}

\begin{figure}[h]
\includegraphics[width=7.5cm]{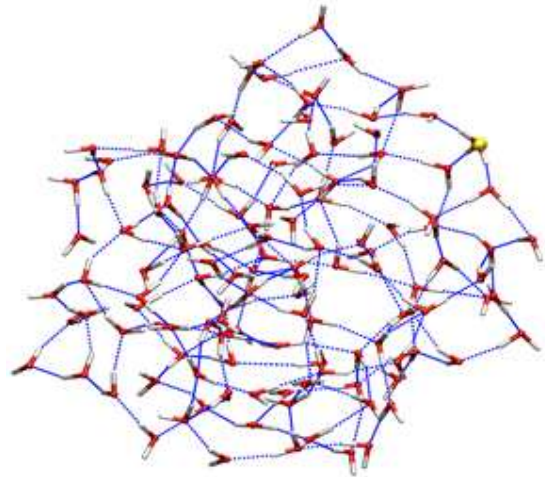}
\caption{\label{fig:cluster_nocon} Cluster configuration from a molecular dynamics run at $T=280$ K without bias. The hydronium oxygen of the EVB-state with the largest weight is depicted in yellow and hydrogen bonds are shown as blue dashed lines. Typically, the excess proton exists as an Eigen-cation at the cluster surface, but other configurations also occur.}
\end{figure}

\begin{figure}[h]
\includegraphics[width=6.5cm]{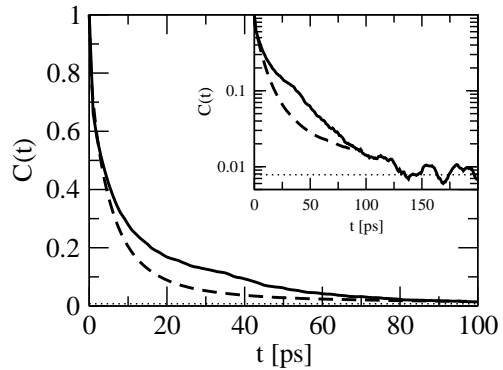}
\caption{\label{fig:C} Time correlation function $C(t)=\langle h(0)h(t)\rangle / \langle h \rangle$ for proton transfer (solid line) and hydrogen bond dynamics (dashed line). 
The horizontal dotted line denotes the long-time limit of the correlation function for proton transfer, $C\rightarrow 1/N_W$. The inset shows the same curves with a logarithmic scale on the $y$-axis.
}
\end{figure}

A typical configuration of the protonated cluster observed in a molecular dynamics simulation carried out at $T=280$ K without bias is shown in Fig. \ref{fig:cluster_nocon}.  Visual inspection of sequences of cluster configurations does not show any indication of crystallinity and indicates that proton transfer and the reorganization of the hydrogen bond network occurs on a picosecond time scale. The qualitative perception that the cluster is in its liquid state can be made quantitative by considering appropriate time correlation functions. For the dynamics of hydrogen bonds,  the time correlation function 
\begin{equation}
C_{\rm HB} (t) = \frac{\langle h_{ij}(0)h_{ij}(t)\rangle}{\langle h_{ij}\rangle}
\end{equation}
measures the conditional probability that a hydrogen bond exists between two particular water molecules at time $t$ provided it existed at time $0$. This time correlation function was introduced by Luzar and Chandler to study the dynamics of hydrogen bonds in bulk water \cite{LUZAR_CHANDLER_PRL,LUZAR_CHANDLER_NATURE}. In the above expression the indicator function $h_{ij}$ is unity if there is a hydrogen bond from molecule $i$ to molecule $j$ and it vanishes otherwise. Here, a hydrogen bond is defined to exist if the $R_{\rm OO}$ distance is less than 3.5 {\AA} and the HOO-angle is less than 30$^\circ$. The behavior of $C_{\rm HB} (t)$ shown in Fig. \ref{fig:C} for the cluster is very similar to that of bulk liquid water at room temperature \cite{LUZAR_CHANDLER_NATURE} indicating that the kinetics of hydrogen bonds in our cluster occurs at approximately the same time scales as that in the liquid phase.

%-----------------------------------------------
%-----------------------------------------------

An analogous time correlation function can be used to quantify the kinetics of proton transfer:
\begin{equation}
C_{\rm p} (t) = \frac{\langle h_{i}(0)h_{i}(t)\rangle}{\langle h_{i}\rangle}.
\end{equation}
The indicator function $h_i(t)$ is unity if oxygen $i$ is the hydronium oxygen in the EVB-state with the largest weight at time $t$ and it vanishes otherwise. The correlation function $C_{\rm p} (t)$ measures the conditional probability that the excess proton localized near oxygen $i$ at time $0$ is still there at a time $t$ later. Similar time correlation functions have been previously used by Vuilleumier and Borgis \cite{BORGIS3} and Schmitt and Voth \cite{VOTH2} to study proton transfer in bulk water. The form of the correlation function $C_{\rm p} (t)$ results from an interplay between the hopping of the excess charge from one molecule to another and the diffusion of water molecules in the cluster. The similarity of the two time correlation functions for hydrogen bonding and proton transfer shown in Fig. \ref{fig:C} is a reflection of the crucial role played by hydrogen bonds in the proton transfer process \cite{AGMON}.

%-----------------------------------------------
%-----------------------------------------------

\begin{figure}[th]
\includegraphics[width=6.5cm]{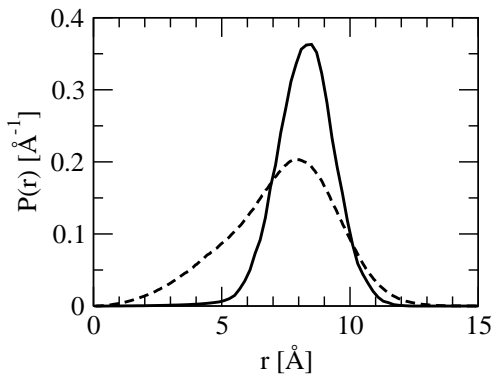}
\caption{\label{fig:Pr} Probability distribution $P(r)$ of the distance $r$ of the center of charge (solid line) and of water oxygens (dashed line) from the center of mass of the cluster.}
\end{figure}

As exemplified by the configuration shown in Fig. \ref{fig:cluster_nocon}, the excess proton is predominantly located at the cluster surface. The stabilization of this position is sufficiently strong to prevent the excess charge from visiting the cluster interior even in molecular dynamics simulations that are several nanoseconds long. This observation is confirmed by free energy calculations carried out with biased simulations as described in Secs. \ref{sec:bias} and \ref{sec:system}. The probability distribution $P_{\rm proton}(r)$ of the distance $r$ of the center of charge from the center of mass, calculated from several biased simulations according to Equ. (\ref{equ:Pbias}), is shown in Fig. \ref{fig:Pr} along with the distribution $P_{\rm water}(r)$ of the distance $r$ of water oxygens from the center of mass obtained from an unbiased simulation. As can be inferred from the figure, the distribution for the proton is peaked at a slightly larger distance than that for the water molecules and falls off more rapidly particularly for smaller distances.

%-----------------------------------------------
%-----------------------------------------------

The form of the probability distributions shown in Fig. \ref{fig:Pr} is strongly influenced by the configuration space available for different values of the radius $r$. To separate this purely geometric factor from less trivial effects we have calculated normalized densities $n(r)$ obtained by comparison with the corresponding distributions for uniform systems.  In particular, the normalized density $n_{\rm water}(r)$ for water molecules was obtained by normalization with the distribution expected in a uniform system with the number density $\rho_{\rm bulk}=55.5 \, {\rm mol}/{\rm l}$ of bulk water, $n_{\rm water}(r)=N P_{\rm water}(r) / 4\pi r^2 \rho_{\rm bulk}$. In other words, $n_{\rm water}(r)$ is the observed number density at distance $r$ from the center of mass compared to the number density of the bulk. In the core of the cluster, i.e., for $r<7-8 \, {\rm \AA}$, the water density is approximately constant and slightly exceeds the density of bulk water. Between 8 {\AA} and 12 {\AA} the water density decreases smoothly to zero. This water density profile indicates that while the cluster fluctuates in shape it remains mostly compact. For the center of charge the normalized density $n_{\rm proton}(r)=P_{\rm proton}(r) / 4\pi r^2 \rho_{\rm uniform}$ is obtained by comparison with a uniform proton density $\rho_{\rm uniform}$ corresponding to one single proton in a sphere of radius $R_0 = 10.8$ {\AA}, which contains 99\% of the proton density. The proton density peaks at about $9$ {\AA}, approximately where the water density begins to decrease from its bulk value. In the cluster core, out to a distance of about 5 {\AA}, the proton density is essentially constant at a value that is smaller than the peak density by a factor of about 40. Note that even though this factor is not exceedingly large, spontaneous excursions of the proton to the cluster center are very rare. One can estimate that during a bias-free MD-simulation with a duration of 1 $\mu$s the 
protonic defect would spend less than 50 ps in the core region closer than
1 {\AA} to the center of mass of the cluster. With our formalism, statistically adequate sampling of all proton positions including the cluster core can be achieved using much shorter simulations.
  
%-----------------------------------------------
\begin{figure}[th]
\includegraphics[width=6.5cm]{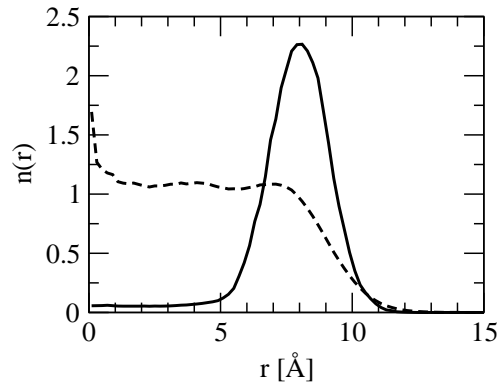}
\caption{\label{fig:rho} Normalized density $n(r)$ of the center of charge (solid line) and of water oxygens (dashed) as a function of the distance $r$ to the center of mass.  
}
\end{figure}
%-----------------------------------------------

\begin{figure}[h]
\includegraphics[width=6.5cm]{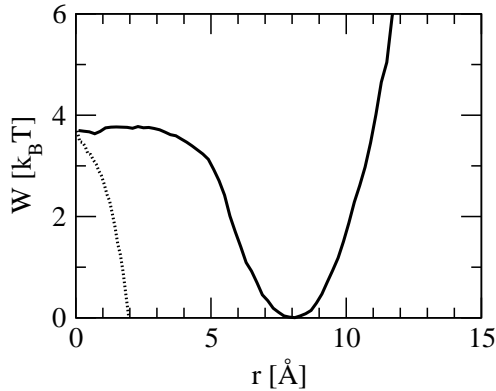}
\caption{\label{fig:PMF} Potential of mean force $W(r)=-k_{\rm B}T \ln n(r)$ as a function of the distance $r$ of the center of charge from the center of mass of the cluster (solid line). The potential of mean force is shifted such that its minimum value is zero. The dotted line indicates the potential of mean force obtained from an incorrect biased MD-simulation, in which the first term on the right hand side of Equ. (\ref{equ:dqdx}) was neglected (see main text).}
\end{figure}

From the normalized proton density $n_{\rm proton}(r)$ the potential of mean force 
\begin{equation}
W(r) = -k_{\rm B}T \ln n_{\rm proton}(r),
\end{equation}
follows. The potential of mean force, which is the reversible work required to change the distance $r$ of the proton defect from the center of mass of the cluster, differs from the free energy $F(r)=-k_{\rm B}T \ln P_{\rm proton}(r)$ by an entropic term that depends on the surface area of a spherical shell with radius $r$. The potential of mean force for our 128-molecule cluster is depicted in Fig. \ref{fig:PMF}, where $W(r)$ was normalized such that it vanishes at its minimum. To carry the proton from the surface of the cluster at $r\approx 8$ {\AA} into its interior a reversible work of about 4 $k_{\rm B}T$ must be expended. Within a core with a radius of  about 4 {\AA} the proton can be translated at no cost. This stabilization of the surface position is very similar in magnitude to the stabilization of an excess proton near a carbon nanotube membrane with respect to the bulk found in recent simulations \cite{PRL_MEMBRANE}. The preferential location of the excess charge near surfaces was noted in earlier simulations \cite{INTERFACE_SIM} and is consistent with experimental data \cite{INTERFACE_EXP}. From the viewpoint of continuum electrostatics this is unexpected because a charge buried deep in the droplet is solvated in a more favorable way. The molecular details of the solvation structure of the excess proton, however, lead to a preferred surface position, presumably due to the strain exerted on the surrounding hydrogen bond network when the excess proton is in the cluster core.

%-----------------------------------------------
%-----------------------------------------------

One may be tempted to simplify the calculation of bias forces by neglecting those force contributions  that arise from the derivatives $\partial c_{0i} /\partial x^\alpha$ of the ground state expansion coefficients $c_{0i}$ on the right hand side of Equ. (\ref{equ:dqdx}). In this approximation, the atomic forces resulting from the bias are simply obtained as weighted average  with weights $c_{0i}^2$ of the bias forces in the individual diabatic states. We have inspected the magnitudes of these two force terms for the bias potentials used in this study and we have found that both forces are of similar magnitude. Neglecting the derivatives  $\partial c_{0i} /\partial x^\alpha$ is therefore unjustified. To exemplify the effect of this approximation we have calculated the potential of mean force $W(r)$ from an incorrect biased simulation with $k_{\rm bias}=2.0$  kcal/mol \AA$^{-2}$ and $R_{\rm bias}=0.0$ {\AA}, in which the first term on the right hand side of Equ. (\ref{equ:dqdx}) was neglected (of course, in such a calculation energy is not conserved even without thermostat). The potential of mean force resulting from this calculation, shown as dotted line in Fig. \ref{fig:PMF}, strongly deviates from the correct $W(r)$. A complete force calculation using the formalism of Sec. \ref{sec:bias} is therefore crucial for a correct free energy calculation.  

%-----------------------------------------------
\begin{figure}[h]
\includegraphics[width=7.5cm]{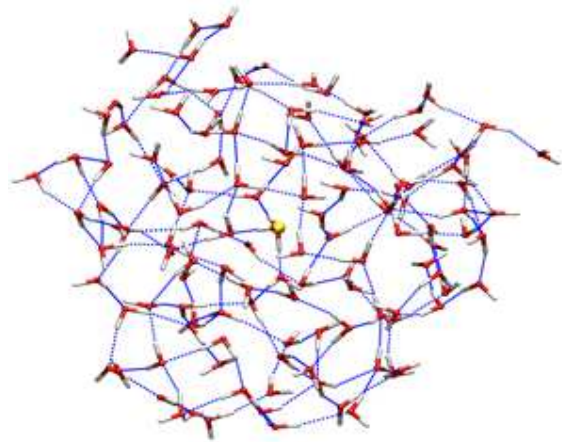}
\caption{\label{fig:cluster_con} Cluster configuration from a molecular 
dynamics simulation in which the center of charge was forced to stay close 
to the center of mass by a bias potential with $R_{\rm bias}=0.0$ {\AA} 
and $k_{\rm bias}=10.0$ kcal/mol \AA$^{-2}$. As in Fig. \ref{fig:cluster_nocon} 
the hydronium oxygen of the EVB-state with the largest weight is depicted 
in yellow. Also in the interior of the cluster the protonic defects exists 
preferentially as Eigen cations.}
\end{figure}
%-----------------------------------------------

It is interesting to compare the solvation structure of the excess proton at the surface and in the core of the cluster. A typical configuration from a biased simulation in which the excess proton is near the center of mass of the cluster is shown in Fig. \ref{fig:cluster_con}. Comparison of configurations with the proton at the surface and in the core reveals that the local solvation structure of the excess proton is similar in these two cases. In both cases the excess proton is preferentially found in the Eigen structure,  observed also in bulk water \cite{MARX}, in which a central hydronium ion donates strong hydrogen bonds to three surrounding water molecules. 

%-----------------------------------------------
\begin{figure}[b]
\includegraphics[width=6.5cm]{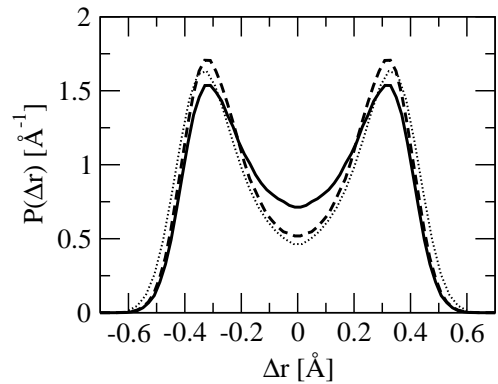}
\caption{\label{fig:dr} Symmetrized distribution for the proton coordinate $\Delta r$ for the unbiased cluster (solid line) and the cluster in which the center of charge was biased to stay close to the center of mass to the cluster (dashed line). Also shown is the distribution of $\Delta r$ in bulk liquid water at $T=300$ K (dotted line) \cite{DELLAGO_CT}.}
\end{figure}
%-----------------------------------------------

This qualitative picture is confirmed by a calculation of the distribution of the proton coordinate $\Delta r$. For the calculation of this coordinate one first determines the hydronium oxygen O$_1$ in the state with the largest weight $c^2_i$. Oxygen O$_2$ is then the oxygen atoms closest to oxygen O$_1$. The proton coordinate $\Delta r$ is calculated from the distances of O$_1$ and O$_2$ to the hydrogen H participating in the hydrogen bond from O$_1$ to O$_2$: $\Delta r = r_{{\rm HO}_2}-r_{{\rm HO}_1}$. Distributions of $\Delta r$ for configurations with the proton at the surface and in the core are shown in Fig. \ref{fig:dr}. The distribution of $\Delta r$ for the core region was calculated from a biased simulation using a bias potential with $R_{\rm bias}=0.0$ {\AA} and $k_{\rm bias}=10.0$ kcal/mol \AA$^{-2}$. In this biased simulation  the center of charge fluctuated around the center of mass with a width of $\langle r^2 \rangle_{\rm bias}=0.1689$ \AA$^2$. As can be inferred from the figure, both asymmetric Eigen-like configurations with $|\Delta r| \approx 0.35$ {\AA} and symmetric Zundel-like configurations with $|\Delta r|\approx 0.0$ {\AA} occur and in both cases, the Eigen-like configurations are preferred. In the core of the cluster, Eigen-like configurations occur slightly more frequently than at the cluster surface. The distribution of $\Delta r$ in the core is very similar to that observed in liquid water at a slightly higher temperature of $T=300$ K shown in Fig. \ref{fig:dr} as a dotted line. Thus, as far as the local structure of the excess proton is concerned, the cluster core provides essentially the same solvating environment as the bulk liquid.

%-------------------------------------------------------------------------
%  CONCLUSIONS
%-------------------------------------------------------------------------

\section{Conclusions}
\label{sec:conclusions}

In the present paper we have presented a new algorithm for the calculation of forces in EVB models where a bias potential is applied to the center of charge, effectively the location of the excess proton. Such bias potentials are necessary if one is interested in exploring unlikely but important configurations that occur, for instance, during transfer events.  The procedure, based on first order perturbation theory, is easy to use  and does not require any significant additional computational effort as all needed quantities are already available from the regular EVB force calculation. Bias forces calculated in this way for the are rigorously correct (for the EVB-model) because they have been derived using perturbation theory rather than by finite differences. 

The algorithm presented in this paper can be used to study the free energetics of proton transfer quantitatively in all systems amenable to an empirical valence bond description including proton transfer in biological systems and acid-base chemistry. As an illustrative example, we have calculated the reversible work required to carry an excess proton from the surface of a small water droplet to its center. The excess charge is preferentially located at the surface where it is stabilized by about 4 $k_{\rm B}T$ with respect to a position in the core of the cluster. This proclivity of the proton to occur near interfaces, which has been previously observed in clusters \cite{VOTH_CLUSTER1,VOTH_CLUSTER2}, at planar liquid-vapor interfaces  \cite{INTERFACE_SIM}, and near hydrophobic membranes \cite{PRL_MEMBRANE}, is  unexpected from continuum electrostatics and is most likely due to the strain exerted by the excess charge on the surrounding hydrogen bond network. Analogous behavior has also been observed for other ions in clusters \cite{WALES,LAAKSONEN,BEYER}.

Another important application of the formalism developed in this paper is the calculation of rate constants for proton translocation for instance through membranes \cite{PRL_MEMBRANE} using the reactive flux formalism of Bennett and Chandler \cite{BENNETT,CHANDLER}, in which two separate simulations are carried out. In both parts the formalism of this paper can be applied. In the first step, one determines the reversible work required to move the proton across the membrane. This can be done efficiently with a molecular dynamics simulation with a bias on the center of charge. In a subsequent step, the so called transmission coefficient is determined by releasing dynamical trajectories from a dividing surface located at the free energy maximum. The initial conditions needed for such a procedure can be generated efficiently by constraining the center of charge to be located on the dividing surface by applying an appropriate bias potential acting on the center of charge. 

%-------------------------------------------------------------------------
%  ACKNOWLEDGMENTS
%-------------------------------------------------------------------------

\section*{Acknowledgments}

The authors acknowledge useful discussions with Gerhard Hummer.
This work was supported by the Austrian Science Fund (FWF) under Grant
No. P17178-N02.  

%-------------------------------------------------------------------------
%  REFERENCES
%-------------------------------------------------------------------------

\end{document}